\documentclass[aip,jap,amsmath,amssymb,reprint]{revtex4-1}

\usepackage[utf8]{inputenc}
\usepackage[T1]{fontenc}
\usepackage{mathptmx}
\usepackage{graphicx}
\usepackage{siunitx}
\usepackage{booktabs}
\usepackage{array}
\usepackage{etoolbox}
\usepackage[colorlinks=true,linkcolor=blue!50!black,citecolor=blue!50!black,urlcolor=blue!50!black]{hyperref}
\usepackage{xcolor}

\makeatletter
\def\@email#1#2{%
 \endgroup
 \patchcmd{\titleblock@produce}
  {\frontmatter@RRAPformat}
  {\frontmatter@RRAPformat{\produce@RRAP{*#1\href{mailto:#2}{#2}}}\frontmatter@RRAPformat}
  {}{}
}%
\makeatother

\newcommand{\Vf}{V_{\mathrm f}}
\newcommand{\Ron}{R_{\mathrm{on}}}
\newcommand{\Roff}{R_{\mathrm{off}}}
\newcommand{\Rs}{R_{\mathrm s}}
\newcommand{\gon}{g_{\mathrm{on}}}
\newcommand{\Pc}{P_{\mathrm c}}
\newcommand{\kf}{k_{\mathrm f}}
\providecommand{\sd}{}\renewcommand{\sd}{\sigma_\delta}
\newcommand{\sw}{\sigma_w}
\newcommand{\PhiN}{\Phi_{\mathcal N}}
\newcommand{\dd}{\mathrm{d}}
\newcommand{\wmin}{w_{\min}}
\newcommand{\phiN}{\varphi_{\mathcal N}}

\begin{document}

\title{Analytical mean-field model for electrochemical metallization memristors with a nanoporous matrix:
application to Ag/anodic alumina/Al}

\author{Lucy A. Valdez}
\affiliation{Departamento de F\'isica, Facultad de Ciencias Exactas y Naturales y Agrimensura (FaCENA),
Universidad Nacional del Nordeste, Av. Libertad 5470, W3404AAS Corrientes, Argentina}
\affiliation{Instituto de Modelado e Innovación Tecnológica, CONICET-UNNE, Av. Libertad 5470, W3404AAS Corrientes, Argentina}

\author{Juan D. Ruiz D\'iaz}
\affiliation{Departamento de Qu\'imica, Facultad de Ciencias Exactas y Naturales y Agrimensura (FaCENA),
Universidad Nacional del Nordeste, Av. Libertad 5470, W3404AAS Corrientes, Argentina}

\author{Rodolfo H. Romero}
\affiliation{Departamento de F\'isica, Facultad de Ciencias Exactas y Naturales y Agrimensura (FaCENA),
Universidad Nacional del Nordeste, Av. Libertad 5470, W3404AAS Corrientes, Argentina}
\affiliation{Instituto de Modelado e Innovación Tecnológica, CONICET-UNNE, Av. Libertad 5470, W3404AAS Corrientes, Argentina}

\author{V\'ictor J. Toranzos}
\email{victoranzos@gmail.com}
\affiliation{Departamento de Ingenier\'ia, Facultad de Ciencias Exactas y Naturales y Agrimensura (FaCENA),
Universidad Nacional del Nordeste, Av. Libertad 5470, W3404AAS Corrientes, Argentina}

\date{\today}

\begin{abstract}
Memristors based on electrochemical metallization with many filaments are usually modeled by microscopic
simulations with many parameters or by phenomenological laws whose parameters are hard to relate to physical
quantities. We present an analytical mean-field model for a population of Ag filaments confined in the
pores of a nanoporous matrix, built on Ohm's law, Faraday's law and a power-based rupture criterion. The
filaments have a truncated normal distribution of cross-sections and rupture leaving lognormally distributed
gaps whose median grows with the dissipated power. Gap closing shifts the distribution of squared gaps rigidly, and the
model therefore admits closed-form expressions for the switching voltage, the read current, the rupture branch and the
high-voltage current. We apply it to Ag/nanoporous anodic alumina/Al structures fabricated from
industrial-grade aluminum, with a \SI{2.4}{\micro m} alumina layer, and measured under ambient conditions.
The parameters were fitted to the bipolar sweeps, with the gap dispersion taken from the temporal shape of the
read after a pulse. They reproduce the bipolar response, including voltage-controlled negative differential
resistance, and predict without refitting the unipolar sweeps and the amplitude of the pulse response, where
state memory appears. A second sample, with a larger contact deposited chemically, is described by the same
model. In vacuum the devices neither switch nor show negative differential resistance, as expected if adsorbed
water supplies the counter reaction; the residual high-voltage current follows ionic hopping conduction.
\end{abstract}

\maketitle
\section{Introduction}

The memristor, proposed by Chua as the fourth passive circuit element~\cite{chua1971}, found a physical
realization when Strukov \emph{et al.} showed that memristance arises naturally in nanoscale systems in which
electronic and ionic transport are coupled by the applied voltage~\cite{strukov2008}. Since then, resistive
memories based on redox processes have become candidates for non-volatile memories and neuromorphic
computing~\cite{waser2007,valov2011,ielmini2016,wang2019}, including proposals that use self-assembled
memristive networks as synapses between electronic neurons~\cite{cipollini2025} and artificial spiking neurons based on Ag diffusive memristors~\cite{midya2025}. Among them, electrochemical
metallization cells switch by the formation and dissolution of filaments of an electrochemically active metal,
typically Ag or Cu, within an insulating matrix. The kinetics of these processes involves nucleation, charge
transfer at the interfaces and ionic migration, and can vary over many orders of magnitude with
voltage~\cite{menzel2013}. \emph{In situ} observations show that metallic inclusions in dielectrics
change their shape, size and position under an electric field, with filament growth modes governed by the
ionic mobility and the redox reaction rates~\cite{yang2014}. The counter reaction at the inert electrode
is easily overlooked; under ambient conditions it is the reduction of adsorbed water, and without it switching
is blocked~\cite{tappertzhofen2013}.

Filamentary switching in oxide films was described long before the memristor revival. In 1970 Dearnaley,
Stoneham and Morgan reviewed forming, voltage-controlled negative differential resistance and memory
phenomena in amorphous films of Al$_2$O$_3$, SiO$_2$ and Ta$_2$O$_5$, and explained them by an ensemble of
conducting filaments that rupture by Joule heating, with a distribution of resistances that determines the
shape of the current--voltage characteristic~\cite{dearnaley1970}. Later models attributed the N-shaped
characteristic of formed structures to a nanometric gap whose width adjusts to the applied
voltage~\cite{mordvintsev2009}; in porous alumina without silver, negative differential resistance of
electronic origin, associated with impurity levels in the oxide, is also observed~\cite{xue2015}. Local
overheating remains a central ingredient of switching in Ag systems, such as Ag$_2$S nanojunctions, whose
switching voltages are explained by a thermally induced phase transition~\cite{gubicza2015}. Alumina is used as the active layer in MIM structures with memristance~\cite{molina2014,huang2020}, and
anodically grown oxides offer a simple, low-cost route to switching layers whose behavior depends strongly on
the electrode used~\cite{leonetti2023}. Anodic alumina films grown in oxalic acid with Ag
contacts show reproducible resistive switching between an insulating and a metallic state, attributed to the
formation and rupture of conducting filaments~\cite{yalishev2009}. All these
systems suffer from a dispersion of the switching parameters, rooted in the random formation and rupture of
filaments. Localizing the switching reduces it, for example with Ag
nanoclusters that concentrate the electric field~\cite{wang2013} or by layers with different defect
concentrations~\cite{huang2020}.

Nanoporous matrices take this localization to the extreme by confining the filaments in channels of defined
geometry. Kolar \emph{et al.} used anodic alumina membranes with Ag nanowires in their pores to down-scale
resistive switching to the nanoscale~\cite{kolar2014}, and in networks of Ag nanowires electrodeposited in
membranes with interconnected pores, memristive behavior, tunneling between nanowires and negative
differential resistance were observed~\cite{chopin2023}. Previous work by our group reported negative
differential resistance with random telegraph noise in oxidized porous silicon devices with Ag
contacts~\cite{marin2015}. Models of many-filament devices tend to sit at two ends. Kinetic Monte Carlo
simulations of filament growth~\cite{menzel2015,dirkmann2015} and switchable resistor networks~\cite{shihong2012}
resolve the microscopic processes at the cost of many parameters, while phenomenological descriptions fit
individual curves with parameters that are difficult to interpret. What is lacking is a description simple
enough to be solved analytically and still able to account, with one parameter set, for the different
electrical responses of the same device.

In this work we study Ag/nanoporous anodic alumina/Al structures measured under ambient conditions and
present an analytical mean-field model for a population of Ag filaments confined in the pores. Rupture follows
Dearnaley's power criterion, and reconnection is governed by the Faradaic closing and dissolution of the gaps
left behind. Because gap closing shifts the gap distribution rigidly, the population can be followed
deterministically, and the switching voltage, the read current, the rupture branch and the high-voltage
current take closed forms. The parameters obtained for the main sample account for its bipolar sweeps,
unipolar sweeps and pulse response, and the model also describes a second sample with a different contact.
Measurements in vacuum, where switching is absent, show the role of the ambient and set an upper bound on
hopping conduction through the alumina.

\section{Methods}

\subsection{Samples and electrical measurements}

\begin{figure*}[t]
\centering
\includegraphics[width=\textwidth]{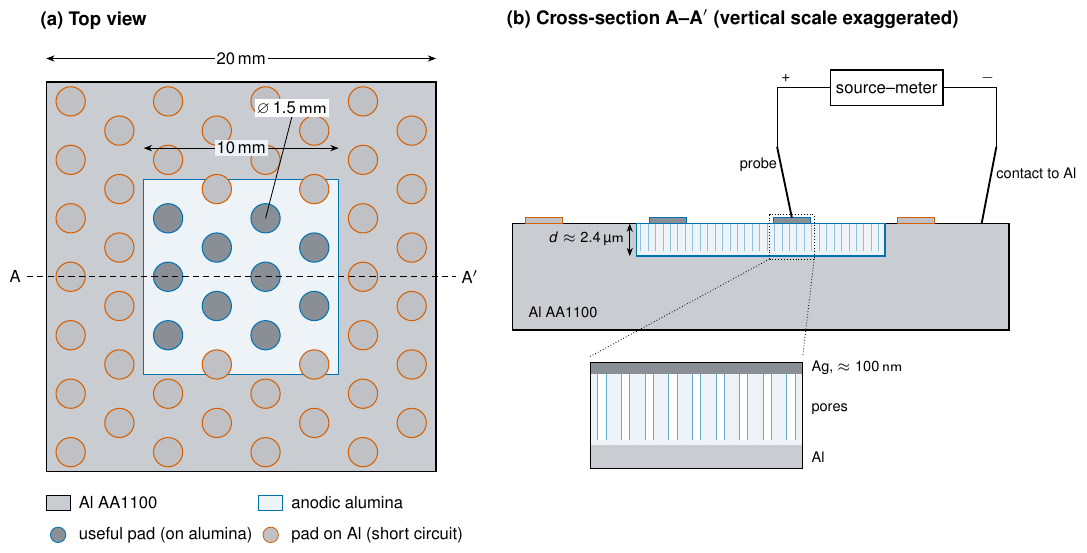}
\caption{Experimental arrangement. (a) Top view of the Al plate with the anodic alumina region and the array
of Ag contacts; only the contacts lying entirely on the alumina (blue edge) are usable, since the rest (orange
edge) are short-circuited to the Al. (b) Cross-section along line A--A$'$, with exaggerated vertical scale, and
connection of the source--meter; the inset shows the pore structure under a contact.}
\label{fig:dispositivo}
\end{figure*}

The metal--insulator--metal (MIM) structures were prepared by anodizing AA1100 aluminum in \SI{0.3}{M}
oxalic acid at a constant current density of \SI{10}{mA/cm^2} for \SI{10}{min}, which yields a nanoporous
alumina layer $\approx\SI{2.4}{\micro m}$ thick (Section~\ref{sec:espesor}). Circular silver contacts about
\SI{100}{nm} thick and \SI{1.5}{mm} in diameter were vacuum-evaporated on the alumina layer, arranged in an
array on the $\SI{20}{} \times \SI{20}{mm}$ Al plate, whose central anodized region measures
$\SI{10}{} \times \SI{10}{mm}$ (Fig.~\ref{fig:dispositivo}). Only the contacts lying entirely on the alumina
form MIM structures; those on the non-anodized aluminum are short-circuited to the bottom electrode and are not
used. After metallization, the samples were annealed for \SI{10}{min} at \SI{200}{\celsius}. The base aluminum
acts as the bottom electrode. Throughout this work the voltage $V$ refers to the Ag contact with respect to the
Al. The analysis below refers to a representative sample; others prepared under the same conditions behaved
similarly, with some dispersion in the parameters. All measurements were
carried out under ambient laboratory conditions, without atmosphere control. Humidity matters in this type of
cell, because the reduction of water adsorbed at the
inert electrode provides the counter reaction that maintains electroneutrality during the oxidation of the
active metal, and its absence blocks switching~\cite{tappertzhofen2013}. Nanoporous anodic alumina retains
adsorbed water in its pores and provides this medium under ambient conditions.

The measurements were made with a home-built programmable voltage source, with an average recording rate of $\approx\SI{0.5}{ms^{-1}}$. 
Three protocols were applied to the main sample:
\begin{enumerate}
\item \textbf{Bipolar sweeps} 
%
%
12 consecutive continuous cycles increasing from 0 to 3 V, then decreasing up to $-3$ V and returning to 0 V, with a voltage step of \SI{12.16}{mV} and sweep rates of 2.8 and \SI{2.3}{V/s} in the increasing and decreasing branches (\SI{4.7}{s} per cycle).
\item \textbf{Unipolar sweeps} $0\to\SI{3}{V}\to0$ (17 consecutive cycles), with the same step (493 points
  per cycle) and an interval of \SI{3.3}{ms} between points (\SI{3.7}{V/s}). The first cycle, with a markedly
  lower current, was regarded as forming and excluded from the averages.
\item \textbf{Pulse sequence} repeated 10 times: \SI{-3}{V} for \SI{1.45}{s}, read at \SI{0.34}{V} for
  \SI{5.8}{s}, \SI{+3}{V} for \SI{0.58}{s} and a new read at \SI{0.34}{V} for \SI{5.8}{s}, sampled every
  \SI{2.9}{ms}. Between segments the device is kept at zero voltage.
\end{enumerate}
The experimental curves are presented as the point-by-point average of the cycles of each protocol, with
error bars equal to one standard deviation between cycles. For the pulses, the 10 sequences were superimposed
from the leading edge of the negative pulse.

For validation, a second MIM structure, fabricated earlier and stored under ambient conditions, was
measured. Its alumina layer was obtained with the same anodizing conditions, but the Ag contact,
\SI{3.5}{mm} in diameter (area $A_2 = \SI{9.62}{mm^2}$, versus $A_1 = \SI{1.77}{mm^2}$ for the main sample),
was chemically deposited by reduction of silver nitrate with Rochelle salt (sodium potassium
tartrate)~\cite{strong1938}. It then received the same \SI{10}{min} annealing at \SI{200}{\celsius}. With the
same system, 10 consecutive bipolar sweeps $0\to\SI{5}{V}\to0$ followed by $0\to\SI{-5}{V}\to0$ were recorded,
with the same voltage step (1648 points per cycle) and the same intervals between points
($\approx\SI{8}{s}$ per cycle). The first cycle, which starts in the ON state because of the previous history
of the device, was excluded from the average.

Throughout this work currents are expressed as the mean current density over the contact, $J = I/A$, with
$A_1 = \SI{1.77}{mm^2}$ for the main sample and the other samples with \SI{1.5}{mm} diameter contacts, and
$A_2 = \SI{9.62}{mm^2}$ for the second sample. Since conduction takes place through the filaments, $J$ is an
average over the contact area, in which the number of active pores is proportional to the area. The model
resistances refer to the whole device.

To assess the role of the ambient, a third sample, prepared in the same way and previously cycled in air,
was measured in vacuum ($\approx\SI{e-2}{mmHg}$, obtained with a mechanical pump). Ten successive manual
up-and-down cycles of the dc voltage between 0 and 9--\SI{10}{V} were applied, in a total of \SI{171}{s}, and
voltage and current were recorded once per second (166 readings; current measured with a multimeter of
\SI{0.01}{\micro A} resolution). Each reading was assigned to the increasing or decreasing branch of its cycle,
and the readings were grouped in \SI{0.5}{V} bins. Readings above \SI{9}{V} and three isolated spikes exceeding
the median of their bin by more than three times its spread (the largest, \SI{4.4}{\micro A}, i.e.\ \SI{2.5}{\micro A/mm^2}, at \SI{8.5}{V})
were discarded. The two branches were averaged separately because they show hysteresis.

\subsection{Alumina layer thickness}
\label{sec:espesor}

\begin{figure}[t]
\centering
\includegraphics[width=\columnwidth]{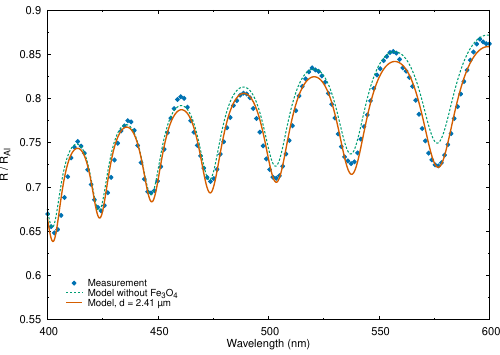}
\caption{Specular reflectance of the alumina layer normalized to that of aluminum (symbols; one point in six
is shown) and fit with the thin-film model (solid line). The dashed line corresponds to the same model without
the Fe$_3$O$_4$ fraction.}
\label{fig:reflectancia}
\end{figure}

The alumina thickness was determined by near-normal-incidence specular reflectance, with an Ocean Optics
HR4000CG-UV-NIR fiber-optic spectrometer (range \SI{200}{} to \SI{1100}{nm}, resolution \SI{0.75}{nm}). The
spectrum, normalized to that of an aluminum surface, was analyzed between \SI{400}{} and \SI{600}{nm}, where
the interference fringes of the film are clearly visible (Fig.~\ref{fig:reflectancia}). The air/alumina/Al
system was modeled with the Airy formula for a thin film, including interface roughness through the Bennett
factor~\cite{bennett1961}. The film was described as an effective medium (Looyenga rule~\cite{looyenga1965})
made of Al$_2$O$_3$, air in the pores and a small fraction of Fe$_3$O$_4$, with the refractive indices of
Malitson for alumina~\cite{malitson1962}, Querry for magnetite~\cite{querry1985} and Raki\'c for
aluminum~\cite{rakic1998}. The fit gives a thickness $d = \SI{2.41}{\micro m}$ ($\mathrm{RMSE} = 0.008$ in
$R/R_{\mathrm{Al}}$), practically independent of the mixing rule (a cylindrical-pore model also gives
\SI{2.41}{\micro m}). The set of solutions with an error up to \SI{10}{\percent} above the minimum spans $d$
between \SI{2.31}{} and \SI{2.51}{\micro m}. Porosity is fully correlated with thickness and remains undetermined
($0.11$--$0.28$). The Fe$_3$O$_4$ fraction is small
($\approx\SI{0.2}{\percent}$ by volume) but necessary, since without it the error doubles; it can be traced to the
iron present in the AA1100 alloy, which is incorporated into the oxide during anodizing.

\subsection{Fuse-filament model}
\label{sec:modelo}

\begin{figure*}[t]
\centering
\includegraphics[width=\textwidth]{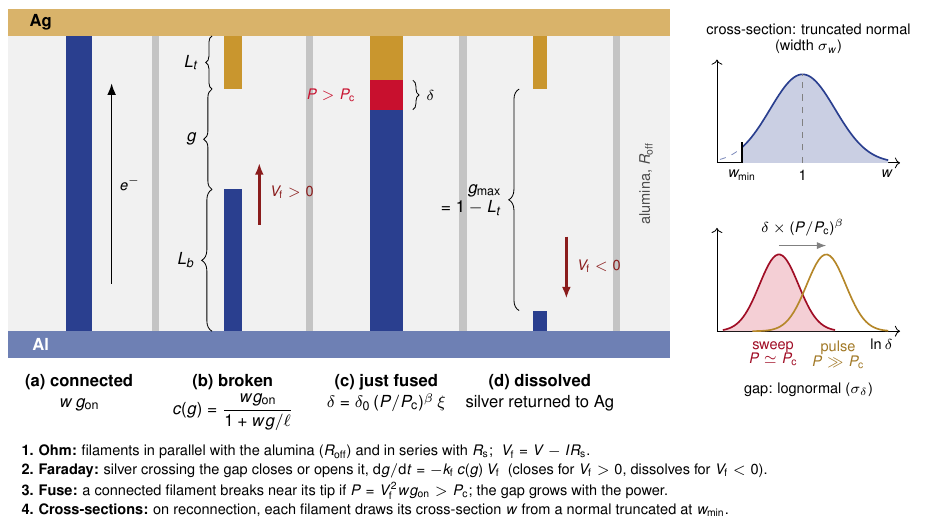}
\caption{Scheme of the model. Each pore contains a filament formed by a segment attached to the Al ($L_b$), a
gap $g$ and a segment attached to the Ag ($L_t$). (a) Connected filament, with conductance $w\gon$. (b) Broken
filament: it conducts through the gap, which closes by silver deposition for $\Vf>0$. (c) Rupture of a
connected filament when the critical power is exceeded, leaving a gap $\delta$ that grows with the dissipated
power. (d) For $\Vf<0$ the segment attached to the Al dissolves up to the maximum gap $1-L_t$. Right: the
distributions of the cross-section (normal truncated at $\wmin$) and of the rupture gap (lognormal, shifted
towards larger gaps when rupture occurs at high power, as during a pulse). The $N$ pores are in parallel with
the alumina conduction ($\Roff$) and the whole is in series with $\Rs$.}
\label{fig:esquema}
\end{figure*}

The device is described as $N$ active pores in parallel, each containing a one-dimensional metallic
filament of length normalized to unity between the Al and the Ag (Fig.~\ref{fig:esquema}). A filament
consists of a segment attached to the Al of length $L_b$, a gap $g$ and a segment attached to the Ag of length
$L_t$, with $g = 1 - L_b - L_t$. Porous anodic alumina has already been used to confine resistive switching at
the nanoscale: Kolar \emph{et al.}~\cite{kolar2014} filled the pores of a membrane with Ag nanowires by
electrodeposition and observed that, with use, silver depletion opens a gap at the top end of the nanowires,
at the interface with the opposite electrode. In networks of Ag nanowires electrodeposited in nanoporous
membranes, the formation and destruction of silver filaments between nanowires also produce memristive
behavior with negative differential resistance in the low-resistance regime~\cite{chopin2023}. The
description of an oxide film as a set of parallel filaments that rupture by Joule heating above a threshold,
with a distribution of resistances that determines the shape of the negative differential resistance, was
proposed by Dearnaley \emph{et al.}~\cite{dearnaley1970} for formed amorphous oxide films. We keep that rupture criterion and add the electrochemical kinetics of gap closing and dissolution, which
controls reconnection. Besides Ohm's law, the model uses only Faraday's law and the power-based rupture
criterion.

\paragraph{Conduction.} A connected filament ($g=0$) has conductance $w\gon$, where $\gon = 1/(N\Ron)$ and
$w$ is a dimensionless cross-section factor. Its distribution is a normal truncated from below,
\begin{equation}
  f(w) \propto \exp\!\left[-\frac{(w-1)^2}{2\sw^2}\right],\qquad w \ge \wmin ,
  \label{eq:fw}
\end{equation}
rescaled so that $\langle w\rangle = 1$. The dispersion $\sw$ sets the width of the distribution and the
minimum cross-section $\wmin$ excludes filaments so thin that they would not rupture within the explored
voltage range. A broken filament conducts through its gap with
\begin{equation}
  c(g) = \frac{w\,\gon}{1 + w g/\ell},
  \label{eq:c}
\end{equation}
where $\ell$ is the conductance of the gap relative to that of the filament. The alumina contributes a
conductance $1/\Roff$ in parallel and the contact a resistance $\Rs$ in series, so that the effective voltage
across the filaments is $\Vf = V - I\Rs$.

\paragraph{Electrochemistry.} The current through a gap transports silver according to Faraday's law. For
$\Vf>0$ the segment attached to the Al grows and the gap closes; for $\Vf<0$ that segment dissolves and the
silver returns to the Ag electrode until it is exhausted ($L_b = 0$):
\begin{equation}
  \frac{\dd g}{\dd t} = -\kf\, c(g)\, \Vf ,\qquad 0 \le g \le 1 - L_t ,
  \label{eq:faraday}
\end{equation}
where $\kf$ is the effective Faraday constant (relative length deposited per unit charge). Connected filaments
have no electrochemical activity.

\paragraph{Rupture.} A connected filament ruptures when the power it dissipates exceeds a critical value,
$P = \Vf^2\, w\, \gon > \Pc$. Rupture occurs near the tip, consistent with the localization of the gap at the
top end of the pores observed in~\cite{kolar2014}: it leaves a segment $L_t$ attached to the Ag (of mean
length $l_t$) and opens a gap that grows with the power dissipated at the moment of rupture,
\begin{equation}
  \delta = \delta_0 \left(\frac{P}{\Pc}\right)^{\beta} \xi ,
  \label{eq:delta}
\end{equation}
where $\xi$ is a lognormal variable with unit median and dispersion $\sd$, and the exponent $\beta$ describes
how much more violent the rupture is when the power exceeds the threshold. When a gap closes again, the
reconnected filament takes a new cross-section factor $w$, drawn from the same distribution, and ruptures
immediately if $P > \Pc$.

Equation~\eqref{eq:delta} separates sweeps from pulses. During a sweep, the voltage rises slowly and each
filament ruptures as soon as it exceeds the threshold ($P \simeq \Pc$): the gaps have median $\delta_0$.
During a high-voltage pulse a filament that reconnects immediately receives a power
$P = w\Vf^2/(\Pc N\Ron)\,\Pc$ far above $\Pc$ and is left with a larger gap. For $\beta = 0$ a gap
distribution independent of the rupture conditions is recovered. A gap whose width grows with the applied
voltage is also the central ingredient of the electroforming model of MIM structures in vacuum by Mordvintsev
\emph{et al.}~\cite{mordvintsev2009}, in which Joule heating regulates the width of a nanometric gap and gives
rise to the N-shaped current--voltage characteristic.

\subsection{Mean-field description}
\label{sec:campomedio}
We call the description mean-field because all filaments are subjected to the same effective voltage
$\Vf = V - I\Rs$, which is in turn determined by the total current $I$ of the population, and spatial correlations
between pores (local potential distribution, current crowding and heating between neighbors) are neglected. Each
filament responds to a common field fixed by the average state of the ensemble, and the population is
described by distributions of cross-sections and gaps rather than filament by filament. In the limit
$\Rs\to0$ the coupling vanishes and the model reduces to independent filaments in parallel, statistically
averaged.

Substituting~\eqref{eq:c} into~\eqref{eq:faraday}, the equation separates:
\begin{equation}
  \frac{\dd}{\dd t}\!\left(\frac{g}{w} + \frac{g^2}{2\ell}\right) = -\kf\,\gon\,\Vf(t).
  \label{eq:A}
\end{equation}
The right-hand side is the same for all broken filaments, because they all share the same $\Vf$. For gaps much
larger than $\ell/w$ (in this work $g/\ell \sim 10^3$) the quadratic term dominates and, in the variable
$s = g^2$,
\begin{equation}
  \frac{\dd s}{\dd t} = -\alpha\,\Vf(t),\qquad \alpha = 2\ell\,\kf\,\gon .
  \label{eq:traslacion}
\end{equation}
The distribution of squared gaps is therefore shifted rigidly by the voltage flux $\Phi(t)=\int \Vf\,\dd t$,
without deforming: an initial gap $g_0$ evolves as $g(t) = [g_0^2 - \alpha\Phi(t)]^{1/2}$. Since, in addition,
$\dd q = c\,\Vf\,\dd t = -\dd g/\kf$, the charge that crosses a gap while it goes from $g_a$ to $g_b$ is exactly
\begin{equation}
  q = (g_a - g_b)/\kf ,
  \label{eq:carga}
\end{equation}
regardless of how the voltage varies in that interval.

With these two properties, individual filaments need not be tracked; it suffices to follow
two populations deterministically: (i) the connected fraction, resolved by cross-section factor $w$, and (ii) the population
of broken filaments, described by its distribution of $s$ and of maximum gap $s_{\max} = (1-L_t)^2$. In each
time interval, the broken population is shifted according to \eqref{eq:traslacion}, the mean ionic current is
obtained exactly from \eqref{eq:carga}, the connected filaments with $w > w^* = \Pc/(\gon \Vf^2)$ move to the
broken population with gaps $\delta$, and those that close their gap reconnect: the fraction with $w \le w^*$
remains connected and the rest ruptures again. At high voltage this closing-and-rupture cycle can repeat many
times within the same interval and is solved exactly. The voltage $\Vf$ is determined self-consistently with
the external circuit ($\Rs$ and $\Roff$). The distribution of $w$ is represented with 3000 equal-probability
classes (1000 during the fit) and that of $\xi$ with 40 classes whose quantiles are shifted at each rupture
according to a low-discrepancy sequence, so that the gaps of successive ruptures are interleaved and the read
current shows no artificial steps. The result is equivalent to the average over infinitely many realizations
of the stochastic filament model, has no sampling noise and does not depend on the time step in the
electrochemical part.

\subsection{Analytical relations}
\label{sec:analiticas}

Several features of the measurements can be written in closed form as functions of a few parameter
combinations. $\PhiN$ and $\phiN$ denote the cumulative normal distribution and its density. The
full derivations are given in Appendix~\ref{ap:deducciones}.

\paragraph{Closing flux.} Integrating~\eqref{eq:A}, a gap $g_0$ closes when the voltage flux reaches
\begin{equation}
  \Phi_c(g_0) = \frac{1}{\kf\gon}\left(\frac{g_0}{w} + \frac{g_0^2}{2\ell}\right) \simeq \frac{g_0^2}{\alpha}.
  \label{eq:phic}
\end{equation}
Switching to ON has no threshold voltage; what matters is the voltage $\times$ time product. In a linear sweep at rate $r$, half of the gaps close at $V_{\mathrm{SET}} = [2r\,\Phi_c(\delta_0)]^{1/2}$,
i.e., the switching voltage scales as $r^{1/2}$.

\paragraph{Read current.} After rupture, a read at constant voltage $V_r$ reconnects the filaments whose gap
satisfies $\delta^2 < \alpha\Phi$. If the gaps have median $\delta_{\mathrm{ef}}$ and lognormal dispersion
$\sd$, for $\Rs \to 0$,
\begin{equation}
  I_{\mathrm{read}}(t) \simeq \frac{V_r}{\Ron}\,\PhiN\!\left[\frac{\tfrac12\ln(\alpha V_r t) - \ln\delta_{\mathrm{ef}}}{\sd}\right] + \frac{V_r}{\Roff}.
  \label{eq:lectura}
\end{equation}
On a logarithmic time scale the rise is a sigmoid of width $2\sd$: going from 10 to 90\,\% of the current
spans a factor $e^{5.13\sd}$ in time. After a pulse of voltage $V_p$, by Eq.~\eqref{eq:delta},
$\delta_{\mathrm{ef}} \simeq \delta_0\,(V_p^2/\Pc N\Ron)^{\beta}$: a stronger pulse leaves larger gaps and a
slower read. A finite series resistance reduces $\Vf$ as the connected fraction grows and makes the rise even
slower.

\paragraph{ON branch during rupture.} Starting with all filaments connected and in the polarity that does not
reconnect them, at each voltage the filaments with cross-section $w < w^* = \Pc N \Ron / \Vf^2$ survive. Since
each filament conducts in proportion to $w$, with the distribution~\eqref{eq:fw} (with parameters $\mu$, $s$ and
$w_m$ once rescaled to unit mean; $z = (w-\mu)/s$),
\begin{equation}
  \begin{split}
  I_{\mathrm{ON}}(\Vf) ={}& \frac{\Vf}{\Ron\,[1-\PhiN(z_m)]}\,
  \bigl\{\mu\,[\PhiN(z^*) - \PhiN(z_m)]\\
  &- s\,[\phiN(z^*) - \phiN(z_m)]\bigr\} + \frac{\Vf}{\Roff}.
  \end{split}
  \label{eq:on}
\end{equation}
The position of the current maximum is set by $\Pc N\Ron$ and its width by $\sw$. Rupture ends at the voltage
$(\Pc N\Ron/w_m)^{1/2}$ at which the thinnest filaments rupture. This expression coincides with the static
characteristic of Dearnaley \emph{et al.}~\cite{dearnaley1970} written for a distribution of filament
resistances $\rho = 1/(w\gon)$; already in that analysis the absence of very high-resistance filaments, which
here corresponds to the minimum cross-section $\wmin$, was necessary for the current to vanish above the
minimum of the characteristic.

\paragraph{High-voltage current.} If every filament that reconnects ruptures again, each one repeats cycles in
which it closes a gap $\delta$ in a time $\delta^2/(\alpha \Vf)$ while carrying the charge $\delta/\kf$. By the
renewal--reward theorem~\cite{ross2014}, the mean current is
\begin{equation}
  I_{\mathrm{HV}} = \frac{2\ell\, e^{-3\sd^2/2}}{\bar\delta}\,\frac{\Vf}{\Ron} + \frac{\Vf}{\Roff},
  \qquad \bar\delta = \delta_0 \left\langle (w/w^*)^{\beta}\right\rangle_{w>w^*},
  \label{eq:cola}
\end{equation}
ohmic and reduced with respect to the ON branch by the factor $2\ell e^{-3\sd^2/2}/\bar\delta$.

\subsection{Fitting procedure}
\label{sec:ajuste}

For both samples the parameters were obtained from the average bipolar curve, the response the model is
required to reproduce quantitatively. For the sweeps, three consecutive cycles
were simulated with the measured waveform and its time intervals, and the last one was compared, so as to reach
the cyclic regime. The bipolar sweep, however, does not determine the parameters that govern the time evolution
of reconnection, the exponent $\beta$ and the gap dispersion $\sd$, because in it ruptures always occur near the
threshold. These parameters were fixed jointly with the temporal shape of the read after the positive pulse,
without fitting its amplitude: the read current increases from sequence to sequence (from \SIrange{2.8}{3.4}{mA/mm^2} in
the first ones to \SIrange{4.5}{5.1}{mA/mm^2} in the intermediate ones), as does the ON conductance of the sweeps over the
cycles, while in the model $\Ron$ is constant. The unipolar sweeps were not used in the fit and served as
validation.

Ten parameters were fitted: $\Ron$, $\Rs$, $\Pc$, $\sw$, $\wmin$, $\kf$, $\delta_0$, $\sd$, $\beta$ and $l_t$.
$N = 200$, $\Roff = \SI{100}{\kilo\ohm}$ and $\ell = 10^{-4}$ were fixed. In the mean-field model $N$ is not
separately identifiable: only the combinations $\Pc N\Ron$ and $\kf/(N\Ron)$ enter. The discretization into
classes introduces small jumps in the cost function, so a derivative-free optimization method was used
(Nelder--Mead simplex, in logarithmic variables), starting from several initial points and keeping the one
with the lowest residual.

\section{Results}
\subsection{Parameters}

Table~\ref{tab:param} summarizes the fitted parameters for the main sample and the most relevant derived
quantities. The total ON resistance is $\Ron + \Rs \approx \SI{24.5}{\ohm}$. A filament of mean cross-section
ruptures at $\Vf = (\Pc N\Ron)^{1/2} \approx \SI{0.74}{V}$ and the thinnest ones of the distribution at
\SI{1.6}{V}. The cross-section distribution has a relative dispersion of 0.53 and its lower cutoff removes
\SI{7}{\percent} of the normal.

The closing kinetics is characterized by three fluxes. The median gap left by a rupture during a sweep closes
with $\Phi_c(\delta_0) \approx \SI{7.6}{V.ms}$, about \SI{22}{ms} at the read voltage. During the \SI{+3}{V}
pulse the rupture power of a mean filament is $\approx(3/0.74)^2\Pc \approx 16\,\Pc$, and by
Eq.~\eqref{eq:delta} the gaps become $16^{\beta} \approx 2.3$ times larger: closing them requires
$\approx\SI{40}{V.ms}$, about \SI{120}{ms} of reading, and the dispersion $\sd \approx 0.85$ spreads
reconnection over more than two orders of magnitude in time. The maximum gap left by dissolution under
negative bias requires $\Phi_c(1-l_t) \approx \SI{3.0e3}{V.ms}$, a value that exceeds the flux of a complete
read ($\SI{0.34}{V}\times\SI{5.8}{s} \approx \SI{2.0e3}{V.ms}$) but is smaller than that provided by the
positive half-cycle of a sweep ($\approx\SI{3.6e3}{V.ms}$). This hierarchy allows the device to remain OFF
during the read after the negative pulse and, instead, to reconnect on the return branch of a sweep.

\begin{table*}[t]
\centering
\caption{Model parameters for the main sample, obtained from the joint fit (bipolar sweep and shape of the
pulse sequence), and derived quantities. $\Pc$ and $\kf$ are given for the nominal scale $N = 200$
($\Pc \propto 1/N$, $\kf \propto N$); the other quantities do not depend on $N$.}
\label{tab:param}
\small
\begin{tabular}{@{}lll@{}}
\toprule
Parameter & Value & Note\\
\midrule
Resistance with all filaments connected, $\Ron$ & \SI{20.0}{\ohm} & fitted\\
Series resistance, $\Rs$ & \SI{4.5}{\ohm} & fitted\\
Critical rupture power, $\Pc$ & \SI{0.137}{mW} & fitted ($N=200$)\\
Cross-section dispersion, $\sw$ & \num{0.53} & fitted\\
Minimum cross-section, $\wmin$ & \num{0.22} & fitted\\
Effective Faraday constant, $\kf$ & \SI{6.66}{mA^{-1}.ms^{-1}} & fitted ($N=200$)\\
Median gap after a rupture at threshold, $\delta_0$ & \num{0.050} & fitted\\
Lognormal gap dispersion, $\sd$ & \num{0.85} & fitted\\
Exponent of the gap with power, $\beta$ & \num{0.30} & fitted\\
Mean length of the segment attached to the Ag, $l_t$ & \num{0.009} & fitted\\
Alumina resistance, $\Roff$ & \SI{100}{\kilo\ohm} & fixed\\
Relative gap conductance, $\ell$ & $10^{-4}$ & fixed\\
\midrule
Rupture voltage of a mean filament, $(\Pc N\Ron)^{1/2}$ & \SI{0.74}{V} & Eq.~\eqref{eq:on}\\
Rupture voltage of the thinnest filaments & \SI{1.6}{V} & Eq.~\eqref{eq:on}\\
Closing flux of the median gap, $\Phi_c(\delta_0)$ & \SI{7.6}{V.ms} & Eq.~\eqref{eq:phic}\\
Closing flux after the \SI{+3}{V} pulse, $\Phi_c(2.3\,\delta_0)$ & \SI{40}{V.ms} & Eqs.~\eqref{eq:phic}, \eqref{eq:delta}\\
Closing flux of the maximum gap, $\Phi_c(1-l_t)$ & \SI{3.0e3}{V.ms} & Eq.~\eqref{eq:phic}\\
\bottomrule
\end{tabular}
\end{table*}

\begin{figure*}[t]
\centering
\includegraphics[width=\textwidth]{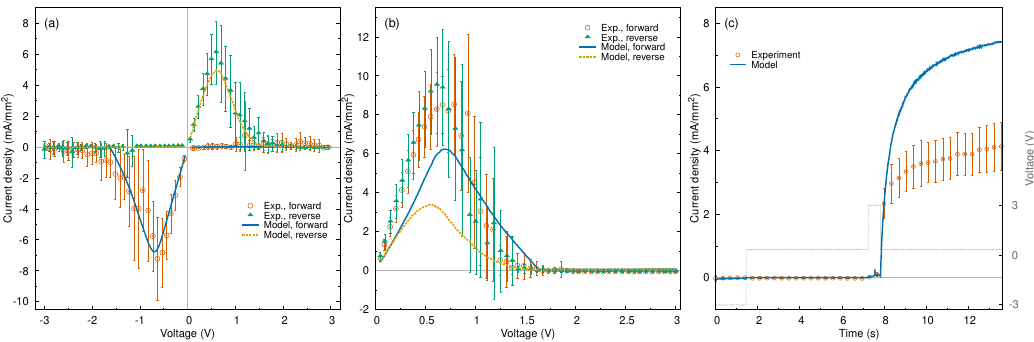}
\caption{Comparison between the mean-field model (lines) and the measurements of the main sample (symbols:
average over cycles; bars: $\pm$ one standard deviation). (a) Bipolar sweeps $0\to3\to-3\to0$~V, $n=12$, used
in the fit. (b) Unipolar sweeps $0\to3\to0$~V, $n=16$, without refitting (validation). In (a) and (b) the
forward ($|V|$ increasing) and return ($|V|$ decreasing) branches are distinguished. (c) Pulse sequence
averaged over $n=10$ repetitions, with the model on an absolute scale. The grey dotted line indicates the applied voltage (right axis).}
\label{fig:resultados}
\end{figure*}

\subsection{Bipolar sweeps}

Figure~\ref{fig:resultados}(a) compares the averaged bipolar sweep with the model; the root-mean-square error
is \SI{0.42}{mA/mm^2}. On the positive forward branch ($0\to\SI{3}{V}$) the device stays OFF. The gaps opened by
dissolution in the previous cycle close only above the rupture voltage, so every filament that reconnects
breaks again at once and the current remains at the value of Eq.~\eqref{eq:cola}, below \SI{0.06}{mA/mm^2}.
On the way back from \SI{3}{V}, filaments that close their gap once $\Vf$ is below threshold stay connected.
The last ruptures happened near threshold and left small gaps, so reconnection is fast and the current
reaches \SI{5.0}{mA/mm^2} at \SI{0.60}{V} (measured: \SI{6.2}{mA/mm^2} at \SI{0.65}{V}) before following the
ohmic branch to the origin. The negative forward branch starts from ON, and the filaments break in order of
decreasing cross-section as described by Eq.~\eqref{eq:on}; the predicted minimum is \SI{-6.7}{mA/mm^2} at
\SI{-0.69}{V} (measured: \SI{-7.4}{mA/mm^2} at \SI{-0.62}{V}), and the last, thinnest filaments break at
\SI{-1.6}{V}. On the negative return, dissolution opens the gaps to their maximum width and the device
remains OFF, which is the starting point of the next cycle.

The amplitude differences at the extrema are within the measured cycle-to-cycle dispersion. The minimum
cross-section $\wmin$ is necessary for the OFF current to be negligible: without it, the thinnest filaments of
the normal would remain connected throughout the cycle and add a background conduction of
$\approx\SI{0.6}{mA/mm^2}$ at $\pm\SI{3}{V}$, which is not observed in this sample.

\subsection{Unipolar sweeps}

With the same parameters, the model reproduces the shape of the unipolar sweeps
[Fig.~\ref{fig:resultados}(b)], with a root-mean-square error of \SI{1.9}{mA/mm^2} between averages. Since there is
no dissolution in this protocol, reconnection during the return leaves the device ON at the start of the next
cycle: the forward branch begins on the ohmic branch and shows a maximum of \SI{6.2}{mA/mm^2} at \SI{0.70}{V}
(measured: \SI{8.5}{mA/mm^2} at \SI{0.79}{V}), followed by rupture. Above \SI{1.7}{V} the current is practically
zero in the model and in the experiment. The main discrepancy is in the return branch: in the measurement it
conducts more than the forward branch (maximum of \SI{9.6}{mA/mm^2} at \SI{0.63}{V}), whereas in the model
reconnection is incomplete at the end of the return (\SI{3.4}{mA/mm^2} at \SI{0.54}{V}) and is completed during the
next forward branch.

\subsection{Pulse response}

Figure~\ref{fig:resultados}(c) shows the averaged pulse sequence. After the negative pulse the device stays
OFF during the whole read (\SI{0.004}{mA/mm^2} in the model, \SI{-0.016}{mA/mm^2} measured). During the
\SI{+3}{V} pulse the filaments close and break repeatedly, and only a small current flows (\SI{0.06}{mA/mm^2}
versus \SI{0.08}{mA/mm^2}). In the read that follows, the current rises slowly and has not saturated after
\SI{5.8}{s}. This slow rise follows of Eq.~\eqref{eq:delta}: ruptures at high power during the pulse leave larger and
widely dispersed gaps, which close throughout the read, whereas in the sweeps ruptures at threshold leave
small gaps that close quickly. If the gap distribution did not depend on power ($\beta = 0$), the fit would have to
favor one behavior over the other; in that case the read saturates halfway through the step.

The largest discrepancy concerns the amplitude of the read. The final current of the model (\SI{7.4}{mA/mm^2}) is 1.8
times the measured one (\SI{4.1}{mA/mm^2}), and the root-mean-square error of the complete sequence is
\SI{1.9}{mA/mm^2}. A difference of scale was to be expected, since only the shape of the read was fitted and $\Ron$ was
determined by the sweeps, measured after the pulses, when the ON conductance had already increased with use
(see Scope and limitations). In the first instants of the read the measured current rises faster than the model
(\SI{2.0}{mA/mm^2} versus \SI{1.6}{mA/mm^2} at \SI{0.1}{s}); from $\approx\SI{0.2}{s}$ onwards the model exceeds it and
concentrates most of its rise within the first second, whereas the measured current grows more evenly throughout
the read.

\subsection{Second sample}
\label{sec:resM2}

\begin{figure}[t]
\centering
\includegraphics[width=\columnwidth]{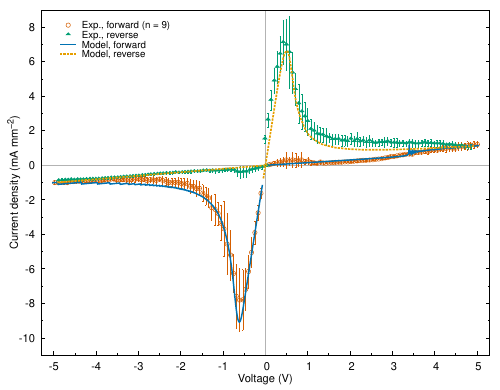}
\caption{Second sample (\SI{3.5}{mm} contact): current density of the bipolar sweeps $0\to5\to-5\to0$~V
($n=9$; symbols: average, bars: $\pm$ one standard deviation) and fit of the mean-field model (lines), with the
same alumina resistance as in the main sample ($\Roff = \SI{100}{\kilo\ohm}$).}
\label{fig:muestra2}
\end{figure}

The second sample was fitted with the same model and procedure, starting from its average bipolar curve, with
the alumina resistance fixed at the same value as in the main sample ($\Roff = \SI{100}{\kilo\ohm}$). Since no
pulses were measured on this sample, the exponent $\beta$, which shows up mainly in the read kinetics, cannot
be determined: the bipolar curve is described without it ($\beta = 0$), and fixing it at the value of the main
sample only increases the error from 0.35 to \SI{0.42}{mA/mm^2}. Figure~\ref{fig:muestra2} shows the result,
with a root-mean-square error of \SI{0.35}{mA/mm^2}. The model reproduces the same sequence of states as in the
main sample, the maxima of the positive return ($6.6$ versus $7.3 \pm \SI{1.4}{mA/mm^2}$ measured, at
\SI{0.50}{V} and \SI{0.47}{V}) and of the negative forward branch ($-9.1$ versus $-8.5 \pm \SI{2.7}{mA/mm^2}$,
both at \SI{-0.62}{V}), the gradual decay of the current after both extrema and the background conduction of
the OFF branches ($1.1$ and $\SI{-1.0}{mA/mm^2}$ at $\pm\SI{5}{V}$, versus $1.2$ and $\SI{-1.0}{mA/mm^2}$
measured).

\begin{table*}[t]
\centering
\caption{Comparison of the parameters of both samples. Resistances are also expressed per unit contact area.
In both samples $\Roff$ was fixed at \SI{100}{\kilo\ohm}; in the second, without pulses, $\sd$ and $\beta$ are
not determined.}
\label{tab:m2}
\small
\begin{tabular}{lcc}
\toprule
 & Main sample & Second sample\\
 & ($A = \SI{1.77}{mm^2}$) & ($A = \SI{9.62}{mm^2}$)\\
\midrule
$\Ron$ & \SI{20.0}{\ohm} (\SI{35}{\ohm\square\milli\metre}) & \SI{1.57}{\ohm} (\SI{15}{\ohm\square\milli\metre})\\
$\Rs$ & \SI{4.5}{\ohm} (\SI{8}{\ohm\square\milli\metre}) & \SI{1.00}{\ohm} (\SI{10}{\ohm\square\milli\metre})\\
$\Roff$ & \SI{100}{\kilo\ohm} (fixed) & \SI{100}{\kilo\ohm} (fixed)\\
Mean rupture voltage, $(\Pc N\Ron)^{1/2}$ & \SI{0.74}{V} & \SI{0.64}{V}\\
$\sw$ & \num{0.53} & \num{0.43}\\
$\wmin$ & \num{0.22} & 0 (no cutoff)\\
$\delta_0$ & \num{0.050} & \num{0.146}\\
$\sd$ & \num{0.85} & not determined\\
$\beta$ & \num{0.30} & not determined (0)\\
\bottomrule
\end{tabular}
\end{table*}

Table~\ref{tab:m2} lists the parameters of the two samples. The cross-section dispersions are similar
($\sw = 0.53$ and 0.43), the series resistance per unit area is comparable and the mean rupture voltage differs
by less than \SI{15}{\percent}. The second sample conducts somewhat more per unit area in ON. Qualitatively, the samples differ in their background
conduction: the second sample conducts $\approx\SI{1}{mA/mm^2}$ at
$\pm\SI{5}{V}$ in the OFF branches of both polarities. With $\Roff$ fixed at the same value as in the main
sample, the model attributes this conduction to the absence of a cutoff in the cross-section distribution
($\wmin = 0$): the thinnest filaments have a rupture voltage higher than the maximum applied, remain connected
throughout the cycle and provide a symmetric ohmic conduction through the channels themselves, without the need
for additional conduction through the alumina. The difference with the main sample ($\wmin = 0.22$) may
come from the contact deposition method, since chemical deposition can produce very thin or incomplete channels
that never rupture.

\subsection{Measurement in vacuum}
\label{sec:vacio}

\begin{figure}[t]
\centering
\includegraphics[width=\columnwidth]{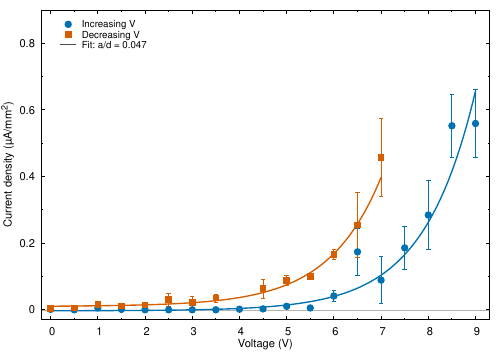}
\caption{Current of a sample previously cycled in air, measured in vacuum ($\approx\SI{e-2}{mmHg}$) during ten
up-and-down voltage cycles. Symbols: average over \SI{0.5}{V} bins for each branch (bars: standard error of the
mean). Lines: fits with the law of field-assisted ionic hopping conduction, Eq.~\eqref{eq:mg}, with the same
$a/d$ in both branches. For the decreasing branch only the fitted range (up to \SI{7}{V}) is shown.}
\label{fig:vacio}
\end{figure}

In vacuum, switching to the low-resistance state does not occur (Fig.~\ref{fig:vacio}). The current is below
\SI{0.01}{\micro A/mm^2} up to \SI{5}{V} and does not exceed $\approx\SI{0.6}{\micro A/mm^2}$ at \SI{9}{V}; that is, the
resistance stays above $\sim\SI{5}{\mega\ohm}$, compared with tens of ohms in the ON state measured in air. The
absence of switching is consistent with the role of humidity noted in the Methods: without water adsorbed in
the pores, the counter reaction at the Al electrode is inhibited and complete filaments cannot
form~\cite{tappertzhofen2013}. The negative differential resistance seen systematically in air is also absent. In porous alumina MIM emitters with Ti/Au electrodes, without silver, voltage-controlled negative
differential resistance is observed even in vacuum and is attributed to impurity levels and oxygen vacancies in
the oxide~\cite{xue2015}. Its absence in our samples indicates that the negative differential resistance
measured in air is not an electronic effect of the alumina but the result of the electrochemical mechanism
mediated by silver and humidity.

A weak hysteresis remains: between 5 and \SI{8}{V} the decreasing branch conducts several times
more than the increasing one (at \SI{7}{V}, \SI{0.46}{\micro A/mm^2} versus \SI{0.09}{\micro A/mm^2} on average), and some
increasing branches show isolated spikes of up to \SI{2.5}{\micro A/mm^2} that are not repeated in the next reading.
Both features are compatible with the incipient growth of conducting paths during the increasing branch,
without forming stable filaments, possibly sustained by residual water at the pressure reached.

Both branches are described by the law of field-assisted ionic hopping conduction (Mott--Gurney
law)~\cite{mott1940,heuer2005,menzel2013},
\begin{equation}
  I = I_b + C\,\sinh\!\left(\frac{q\,a\,V}{2\,k_BT\,d}\right),
  \label{eq:mg}
\end{equation}
where $a$ is the hopping distance, $d \approx \SI{2.4}{\micro m}$ the alumina thickness
(Section~\ref{sec:espesor}) and $I_b$ a background current, compatible with zero within the resolution. The
weighted least-squares fit to the averages (with $k_BT/q = \SI{26}{mV}$) gives $a/d = 0.047 \pm 0.007$ for the
increasing branch (0--\SI{9}{V}) and $a/d = 0.047 \pm 0.010$ for the decreasing branch (0--\SI{7}{V}); above
\SI{7}{V} the decreasing branch departs from the exponential and tends to a plateau of
$\approx\SI{0.45}{\micro A/mm^2}$ (not shown in the figure). A previous measurement in another series, with a single increasing sweep up to
\SI{11.5}{V}, gave the same value, $a/d = 0.047 \pm 0.007$. The two branches differ only in the amplitude $C$,
about four times larger in the decreasing branch: the transport mechanism is the same, with more active paths
after the increasing branch. Extrapolated to low voltages, this conduction contributes between
$\sim\SI{0.003}{\micro A/mm^2}$ (increasing) and $\sim\SI{0.011}{\micro A/mm^2}$ (decreasing) at \SI{3}{V}, resistances of
$10^8$ to $\SI{e9}{\ohm}$: at the operating voltages (up to $\pm\SI{3}{V}$) hopping conduction through the
alumina is negligible compared with that of the filaments, and the current measured in air corresponds to the
ohmic, drift conduction of the metallic filaments. This justifies large values of $\Roff$ and of the resistance
of open gaps ($\ell \ll 1$) in the model.

Fowler--Nordheim tunneling through the oxide is unlikely to explain the high-voltage conduction. In porous
alumina films about \SI{100}{nm} thick this mechanism dominates emission in vacuum for fields close to
\SI{1}{MV/cm}~\cite{xue2015}; in our \SI{2.4}{\micro m} films, \SI{9}{V} corresponds to a mean field of only
$\approx\SI{0.04}{MV/cm}$, about twenty-five times lower, which favors the hopping interpretation.

The ratio $a/d \approx 0.05$ implies an apparent hopping distance of $\approx\SI{113}{nm}$, much larger than the
site spacing of an ionic conductor, which is of the order of interatomic distances~\cite{menzel2013}. It is close, however, to the median gap left by a rupture in the main sample ($\delta_0 = 0.050$, in units of
the pore length), which suggests that in a cycled sample the high-voltage conduction in vacuum occurs through
the residual gaps of broken filaments, which concentrate the voltage drop. In Ag nanowire networks with gaps
between nanowires, a high-resistance conduction through the gaps that becomes strongly nonlinear at high
voltage has been observed in an analogous way, attributed there to direct tunneling and field
emission~\cite{chopin2023}. This identification is tentative: the hopping distance obtained by fitting
Eq.~\eqref{eq:mg} is in general an apparent quantity, which in disordered conductors can greatly exceed the real
distance and depends on sample thickness~\cite{heuer2005}, and $\delta_0$ is correlated with other model
parameters.

\subsection{Relation to previous observations}
\label{sec:previos}

Features compatible with this mechanism were reported in sandwich-type devices of oxidized porous silicon with
chemically deposited, annealed Ag contacts, measured in air~\cite{marin2015}. After several forming cycles a
voltage-controlled negative differential resistance appeared above \SI{3}{V}; the current at constant voltage
showed random telegraph noise; and on small electrodes (\SI{0.2}{mm^2}) the curves showed discrete resistance
states, which averaged out on electrodes 140 times larger. In that work the effect was attributed to the
blocking of conduction channels by carrier trapping. If the oxidized skeleton of porous silicon plays the role of the alumina, as a matrix for Ag filaments, the
present model gives another reading of those results. The negative differential resistance would then be the
power-driven rupture of filaments as the voltage rises [Eq.~\eqref{eq:on}], as on the forward branch of the
unipolar sweeps in Fig.~\ref{fig:resultados}(b). The telegraph noise would come from the closing-and-rupture
cycles of single filaments, whose average is the current of Eq.~\eqref{eq:cola}. Small electrodes contain few
filaments and therefore show discrete states, which average out on large ones because the number of
filaments scales with the area. The effect reported in the same work for coplanar Al/porous silicon/Al devices, without a silver source, which
appeared in vacuum and disappeared in air, has a different origin and lies outside this
mechanism.

\subsection{Scope and limitations}

No switching threshold is imposed in the model; the voltages at which the device changes state follow
from the rupture power and the closing flux. Compared with a description using lognormal cross-sections and
gaps independent of the rupture conditions, the model adds two ingredients, each with a visible effect on the data. The truncated normal cross-section distribution sets
the width of the current maximum and the subsequent gradual decay (and, without cutoff, the background
conduction), and the dependence of the gap on power makes reconnection fast in the sweeps and slow and sustained
in the read after a pulse.

Most of the remaining discrepancies share one origin: the model assumes an invariant ON conductance, whereas the
measurements show that it increases with use, both cycle to cycle in the sweeps and sequence to sequence in the
pulses. Since the sweeps were measured after the pulses, fitting $\Ron$ to the sweeps overestimates the read
current of the pulses, whose scale was not fitted. Including the thickening of connected filaments with the
transported charge is the natural extension of the model. The counter reaction at the Al electrode, supplied by adsorbed water under ambient
conditions~\cite{tappertzhofen2013}, is not modeled explicitly. Its effect is absorbed into the effective
Faraday constant $\kf$, which may therefore depend on humidity. We also neglect the electromotive force associated with the non-equilibrium states generated by these reactions (up to
\SI{0.6}{V} in SiO$_2$ cells~\cite{tappertzhofen2013}). The number of active filaments, which does not enter the
mean behavior, can be estimated from the cycle-to-cycle dispersion through stochastic realizations of the same
model. The linear Faraday law with an effective constant $\kf$ condenses into a single parameter the nucleation,
interfacial charge transfer and ionic migration, which in detailed models of electrochemical metallization cells
give the switching kinetics a strongly nonlinear voltage dependence~\cite{menzel2013}; for the voltages and
times explored here, a description in terms of the voltage flux was enough. Electrode kinetics, field-assisted
growth, silver diffusion, distributed Joule heating and the interaction between neighboring pores are outside
the present description~\cite{waser2007,valov2011,ielmini2016}.

\section{Conclusions}

We studied Al/nanoporous anodic alumina/Ag structures made from industrial-grade aluminum, with a
$\approx\SI{2.4}{\micro m}$ alumina layer measured by reflectance. Under ambient conditions they behave as
electrochemical metallization memristors in which the pores confine Ag filaments, with bipolar and unipolar
switching, voltage-controlled negative differential resistance and a state memory that shows up in the read
current after a pulse.

A fuse-filament model treated in the mean-field approximation, with Ohm's law, Faraday's law and a
power-based rupture criterion, accounts for the bipolar sweeps, unipolar sweeps and pulse response of the main
sample with one parameter set, and gives closed forms for the switching voltage, the read current, the rupture
branch and the high-voltage current. Sweeps and pulses can be described together only if the cross-sections
follow a truncated normal distribution and the rupture gap grows with the dissipated power. The first
ingredient fixes the width of the current maximum, its slow decay and the background conduction; the second
makes reconnection fast after a sweep and slow after a pulse. In a second sample with a chemically deposited
contact, the background conduction is explained by thin filaments that never fuse, with no extra conduction
through the alumina.

In vacuum the devices neither switch nor show negative differential resistance, which points to adsorbed
water as the counter reaction. The remaining high-voltage current follows field-assisted ionic hopping with
the same $a/d$ on both branches, is negligible at the operating voltages, and gives an apparent hopping
distance close to the median gap of the model, an identification we regard as tentative. The same mechanism
also offers a common explanation for earlier observations in oxidized porous silicon with Ag contacts.

Letting the filaments thicken with the transported charge should account for the increase of conductance with
use, which causes most of the remaining discrepancies, and stochastic realizations of the model would give the
number of active filaments from the cycle-to-cycle dispersion. Circuits that drive the device dynamically,
such as relaxation oscillators of the kind used for artificial neurons with Ag memristors~\cite{midya2025}, and
measurements at controlled humidity would test the model further. Since the fabrication and the model
are simple, these structures are a convenient platform for studying resistive switching and its use in
memories and neuromorphic computing.

\appendix
\section{Derivation of the analytical relations}
\label{ap:deducciones}

\paragraph{A.1. Shift of the gap population.} Combining the conductance of a gap, Eq.~\eqref{eq:c}, with
Faraday's law, Eq.~\eqref{eq:faraday},
\begin{equation}
  \frac{\dd g}{\dd t} = -\kf\,\frac{w\gon}{1+wg/\ell}\,\Vf
  \quad\Longrightarrow\quad
  \left(\frac{1}{w}+\frac{g}{\ell}\right)\dd g = -\kf\gon\,\Vf\,\dd t .
\end{equation}
The left-hand side depends only on $g$ and the right-hand side only on $t$. Integrating,
\begin{equation}
  \frac{g}{w}+\frac{g^2}{2\ell} = \frac{g_0}{w}+\frac{g_0^2}{2\ell} - \kf\gon\,\Phi(t),
  \qquad \Phi(t)=\int_0^t \Vf\,\dd t' .
  \label{eq:apA}
\end{equation}
If $g \gg \ell/w$, the linear term is negligible compared with the quadratic one (its relative weight is
$2\ell/(wg)$, of order $10^{-3}$ in this work). Multiplying by $2\ell$ gives $g^2 = g_0^2 - \alpha\Phi$, with
$\alpha = 2\ell\kf\gon$, which is Eq.~\eqref{eq:traslacion}. Since $\alpha\Phi$ is the same for all broken
filaments, the distribution of $s=g^2$ shifts without changing shape.

\paragraph{A.2. Transported charge.} The charge crossing a gap in $\dd t$ is $\dd q = c(g)\Vf\,\dd t$. By
Eq.~\eqref{eq:faraday}, $c(g)\Vf\,\dd t = -\dd g/\kf$, so that
\begin{equation}
  q = \int \dd q = -\frac{1}{\kf}\int_{g_a}^{g_b}\dd g = \frac{g_a-g_b}{\kf}.
\end{equation}
The result is exact: it requires neither the approximation of A.1 nor knowledge of the form of $\Vf(t)$.

\paragraph{A.3. Closing flux and switching voltage.} Setting $g=0$ in Eq.~\eqref{eq:apA} gives the flux needed
to close a gap $g_0$:
\begin{equation}
  \Phi_c(g_0) = \frac{1}{\kf\gon}\left(\frac{g_0}{w}+\frac{g_0^2}{2\ell}\right) \simeq \frac{g_0^2}{\alpha}.
\end{equation}
In a linear sweep $\Vf = rt$ the accumulated flux is $\Phi = rt^2/2 = \Vf^2/(2r)$. Equating it to
$\Phi_c(\delta_0)$, the median gap closes (half of the filaments are connected) at
\begin{equation}
  V_{\mathrm{SET}} = \sqrt{2r\,\Phi_c(\delta_0)} \simeq \delta_0\sqrt{\frac{r}{\ell\kf\gon}} .
\end{equation}

\paragraph{A.4. Read current.} Immediately after rupture, each filament has a gap $\delta$. During a read at
constant voltage $V_r$ (with $\Rs\to0$, $\Vf = V_r$), the flux is $\Phi = V_r t$ and, by A.1, a gap $\delta$ is
closed if $\delta^2 - \alpha V_r t \le 0$, i.e., if
\begin{equation}
  \delta \le g^*(t) = \sqrt{\alpha V_r t}.
\end{equation}
The connected fraction is the probability $P(\delta \le g^*)$. If $\delta$ is lognormal with median
$\delta_{\mathrm{ef}}$ and dispersion $\sd$, the variable $\ln\delta$ is normal with mean $\ln\delta_{\mathrm{ef}}$
and standard deviation $\sd$. Then
\begin{equation}
  \begin{split}
  P(\delta\le g^*) &= P(\ln\delta \le \ln g^*) = \PhiN\!\left(\frac{\ln g^* - \ln\delta_{\mathrm{ef}}}{\sd}\right),\\
  \ln g^* &= \tfrac12\ln(\alpha V_r t).
  \end{split}
\end{equation}
The factor $\tfrac12$ arises because closing is linear in $g^2$ and not in $g$. The connected filaments conduct
$V_r/\Ron$ in total (the mean cross-section is unity and, at the read voltage, none ruptures), and the alumina adds
$V_r/\Roff$. This gives Eq.~\eqref{eq:lectura}.

The argument of $\PhiN$ is linear in $\ln t$ with slope $1/(2\sd)$. The rise is therefore a sigmoid on a
logarithmic time scale with standard deviation $2\sd$ in $\ln t$, centered at
$t_{1/2} = \delta_{\mathrm{ef}}^2/(\alpha V_r)$. Since 10\,\% and 90\,\% of $\PhiN$ correspond to $\mp1.28$
standard deviations, the interval between them spans a factor $\exp(2\times1.28\times2\sd) = e^{5.13\sd}$ in time.
The lognormal distribution is a model hypothesis, appropriate for quantities resulting from multiplicative
processes; with another distribution, Eq.~\eqref{eq:lectura} still holds with $\PhiN$ replaced by the
corresponding cumulative distribution evaluated at $g^*(t)$. According to Eq.~\eqref{eq:delta}, the gaps left by
a pulse of voltage $V_p$ have median $\delta_{\mathrm{ef}} = \delta_0\langle (P/\Pc)^{\beta}\rangle$, with
$P/\Pc = w V_p^2/(\Pc N\Ron)$; for a filament of mean cross-section,
$\delta_{\mathrm{ef}} \simeq \delta_0 (V_p^2/\Pc N\Ron)^{\beta}$.

\paragraph{A.5. ON branch during rupture.} Let $f(w)$ be the cross-section density, Eq.~\eqref{eq:fw}, a normal
of mean $\mu$ and standard deviation $s$ truncated at $w_m$, with the parameters already rescaled so that
$\langle w\rangle = 1$:
\begin{equation}
  f(w) = \frac{\phiN(z)}{s\,[1-\PhiN(z_m)]},\qquad z = \frac{w-\mu}{s},\quad w \ge w_m .
\end{equation}
A connected filament conducts $w\gon\Vf$ and ruptures if $w > w^* = \Pc/(\gon\Vf^2) = \Pc N\Ron/\Vf^2$. With all
filaments initially connected and in the polarity that does not reconnect them, the current is
\begin{equation}
  \begin{split}
  I_{\mathrm{ON}} &= N\gon\Vf \int_{w_m}^{w^*} w\,f(w)\,\dd w + \frac{\Vf}{\Roff}\\
                 &= \frac{\Vf}{\Ron}\int_{w_m}^{w^*} w\,f(w)\,\dd w + \frac{\Vf}{\Roff}.
  \end{split}
\end{equation}
Writing $w = \mu + s z$ and using $\int z\,\phiN(z)\,\dd z = -\phiN(z)$,
\begin{equation}
  \begin{split}
  &\int_{w_m}^{w^*} w\,f(w)\,\dd w \\
  &\quad= \frac{\mu\,[\PhiN(z^*)-\PhiN(z_m)] - s\,[\phiN(z^*)-\phiN(z_m)]}{1-\PhiN(z_m)},
  \end{split}
\end{equation}
which leads to Eq.~\eqref{eq:on}. For $w^* \to\infty$ the ratio equals $\langle w\rangle = 1$ and the ohmic branch
$\Vf/\Ron$ is recovered. The current is maximum near $w^*\simeq1$, i.e., at $\Vf\simeq(\Pc N\Ron)^{1/2}$, and
vanishes when $w^* = w_m$: rupture is complete at $\Vf = (\Pc N\Ron/w_m)^{1/2}$. Compared with a lognormal of the
same width, the normal has fewer very thick filaments, which would rupture at low voltages, and more thin
filaments, which extend the current decay towards high voltages.

\paragraph{A.6. High-voltage current.} When $w^*$ is smaller than the cross-section of all filaments, each
filament repeats the same cycle indefinitely: it ruptures leaving a gap $\delta$, closes it and ruptures again
instantly. By A.1 and A.2, a cycle with gap $\delta$ lasts $\tau = \delta^2/(\alpha\Vf)$ and transports the charge
$q = \delta/\kf$. The gaps of successive cycles are independent, so the succession of cycles is a renewal process.
By the renewal--reward theorem~\cite{ross2014}, the mean current of a filament is the ratio of the mean values:
\begin{equation}
  \bar\imath = \frac{\langle q\rangle}{\langle\tau\rangle}
            = \frac{\langle\delta\rangle/\kf}{\langle\delta^2\rangle/(\alpha\Vf)}
            = 2\ell\gon\Vf\,\frac{\langle\delta\rangle}{\langle\delta^2\rangle}.
\end{equation}
For a lognormal of median $\bar\delta$, $\langle\delta^n\rangle = \bar\delta^n\, e^{n^2\sd^2/2}$, i.e.,
$\langle\delta\rangle = \bar\delta e^{\sd^2/2}$ and $\langle\delta^2\rangle = \bar\delta^2 e^{2\sd^2}$. In this regime, a
reconnected filament of cross-section $w$ ruptures with $P/\Pc = w/w^*$, so that, by Eq.~\eqref{eq:delta},
$\bar\delta = \delta_0\langle (w/w^*)^{\beta}\rangle_{w>w^*}$ (the correction for the correlation between $w$ and
$\delta$ is second order in $\beta$). Multiplying by the $N$ filaments and adding the alumina conduction,
\begin{equation}
  I_{\mathrm{HV}} = N\bar\imath + \frac{\Vf}{\Roff}
  = \frac{2\ell\,e^{-3\sd^2/2}}{\bar\delta}\,\frac{\Vf}{\Ron} + \frac{\Vf}{\Roff},
\end{equation}
which is Eq.~\eqref{eq:cola}. The rupture time is neglected compared with the closing time.

\begin{acknowledgments}
The authors thank FaCENA--UNNE for funding the projects ``Elaboración y caracterización de dispositivos MIM de alúmina nanoporosa'' (RES-2025-611-D-EXA\#UNNE) and ``Elaboración, caracterización y modelado de Memristores MIM'' (RES-2026-529-D-EXA\#UNNE).
\end{acknowledgments}

\section*{Author declarations}
\subsection*{Conflict of interest}
The authors have no conflicts to disclose.

\subsection*{Author contributions}

\textbf{Lucy A. Valdez:} Investigation; Methodology; Data curation; Resources.
\textbf{Juan D. Ruiz D\'iaz:} Investigation; Methodology; Resources.
\textbf{Rodolfo H. Romero:} Investigation; Methodology; Writing – review \& editing; Formal analysis; Funding acquisition.
\textbf{V\'ictor J. Toranzos:} Conceptualization; Methodology; Writing – review \& editing; Formal analysis; Project administration; Funding acquisition.

\section*{Data availability}
The data that support the findings of this study are available from the corresponding author upon reasonable
request.

\bibliography{references}

\end{document}